\documentclass{IEEEtran}
\usepackage{cite}
\usepackage{amsmath,amssymb,amsfonts}
\usepackage{graphicx}
\usepackage{textcomp,nicefrac}
\usepackage{cuted}
\def\BibTeX{{\rm B\kern-.05em{\sc i\kern-.025em b}\kern-.08em
T\kern-.1667em\lower.7ex\hbox{E}\kern-.125emX}}

\begin{document}
\title{Particle Tracking with Space Charge Effects using Graphics Processing Unit }
\author{Yoshinori Kurimoto
\thanks{We gratefully acknowledge the support of NVIDIA Corporation with the donation of the Quadro P6000 used for this research. We also thank to Susumu Igarashi (KEK) and Takaaki Yasui (The University of Tokyo) for the benchmark with SCTR. Yoshinori Kurimoto is with High Energy Accelerator Research Organization, Tokai-mura, Naka-gun, Ibraki, 319-1195 Japan (e-mail: kurimoto@post.j-parc.jp)}}

\maketitle

\begin{abstract}
Particle tracking simulations with space charge effects are very
important for high-intensity proton rings. Since they include not only
Hamilton mechanics of a single particle but constructing charge
densities and solving Poisson equations to obtain the electromagnetic
field due to the space charge, they are extremely time-consuming. We
have newly developed a particle tracking simulation code that can be
used in Graphics Processing Units (GPU). GPUs have strong capacities
of parallel processing so that the calculation of single-particle
mechanics can be done very fast by complete parallelization. Our new
code also includes the space charge effect. It must construct charge
densities, which cannot be completely parallelized. For the charge
density construction, we can use “shared memory” which can be accessed
very fast from each thread. The usage of shared memory is another
advantage of GPU computing. As a result of our new development, we
increase the speed of our particle tracking including space charge
effect approximately 10 times faster than that in the case of our conventional
code used in CPU.
\end{abstract}

\section{Introduction}
\label{sec:introduction}
\IEEEPARstart{S}{pace} charge effects limit the number of particles that can be formed into a bunch in proton accelerators. Electromagnetic fields in a bunch force individual protons to the outside of the bunch. 
For ring accelerators such as synchrotrons, these defocusing effects shift a betatron tune, which is defined as the number of transverse oscillations per one turn. The betatron tune must be precisely controlled by quadrupole magnets to prevent errors and nonlinear terms of magnetic fields from exciting the transverse oscillations, otherwise space charge effects can cause emittance growth and consequent beam losses. Electromagnetic potentials due to space charge effects generally add nonlinear terms to Hamiltonian of the transverse motion so that the tune shifts of individual protons depends on their transverse amplitude. These different betatron tunes in a bunch can be hardly measured and controlled.  We thus rely on numerical simulation for space charge effects. 

Particle-in-Cell (PIC) methods\cite{Hockney} are widely used for numerical calculation of space charge effects. In PIC methods, the following processes are repeated after 
simulated spaces are divided into many cells with their representative points called grids. 
\begin{enumerate}
\item Charge densities are calculated by assigning each particle to the adjacent grids. 
\item Potentials and fields are obtained by solving the Poisson equations
\item Particles are tracked by solving the equations of motion in the obtained electromagnetic fields
\end{enumerate}
Particles in ring accelerators usually pass thousands of components for a single turn and circulate over thousands of turns. This involves millions of charge densities and potentials to be calculated. In addition, a large number of particles ($10^{5}-10^{6}$) must be simulated for sufficient accuracy.  For these reasons, the PIC methods for ring accelerators require large computational resources. 

We developed a new PIC simulation code executable by graphic processing units (GPUs). Their high parallel computing performance makes multi-particle trackings efficient. Even calculations of the charge densities and potentials,  
which are not completely parallelized, can be accelerated using on-chip shared memory provided by GPUs. Although there are several tracking simulation using GPU that are reported\cite{Soliday:2012zz}\cite{GPUSIM2}\cite{GPUSIM3}, we concentrate on the application for long particle bunches so that two-dimensional PIC simulation can be used. In case of a 100$\times$100 grid for example, all grid cells can be allocated in the on-chip shared memory. This can drastically shorten the time required for making charge distributions. In this paper, the details of our new PIC simulation code are described. 

The paper is organized as follows. In Section~\ref{sec:GPU}, general features of GPUs are briefly described. 
In Section~\ref{sec:SPM}, the treatments of single-particle dynamics in the code are explained. These are about 
particle motions in external fields of the accelerator components such as electromagnets. 
Section~\ref{sec:SCE} describes how the code involves the calculations about space charge effects. This 
is the main part of this article. In Section~\ref{sec:JPARC}, we show some results of the simulation for the J-PARC (Japan Proton Accelerator Research Complex) Main Ring\cite{Koseki} using the code. They are compared with the results by another existing simulator. The speed of the new code is also compared with a similar PIC code running on CPUs.    
Finally, we will summarize this paper in Section \ref{sec:summary}.

\section{Graphic Processing Unit}
\label{sec:GPU}
\begin{figure}[htbp]
\centerline{\includegraphics[width=3.5in]{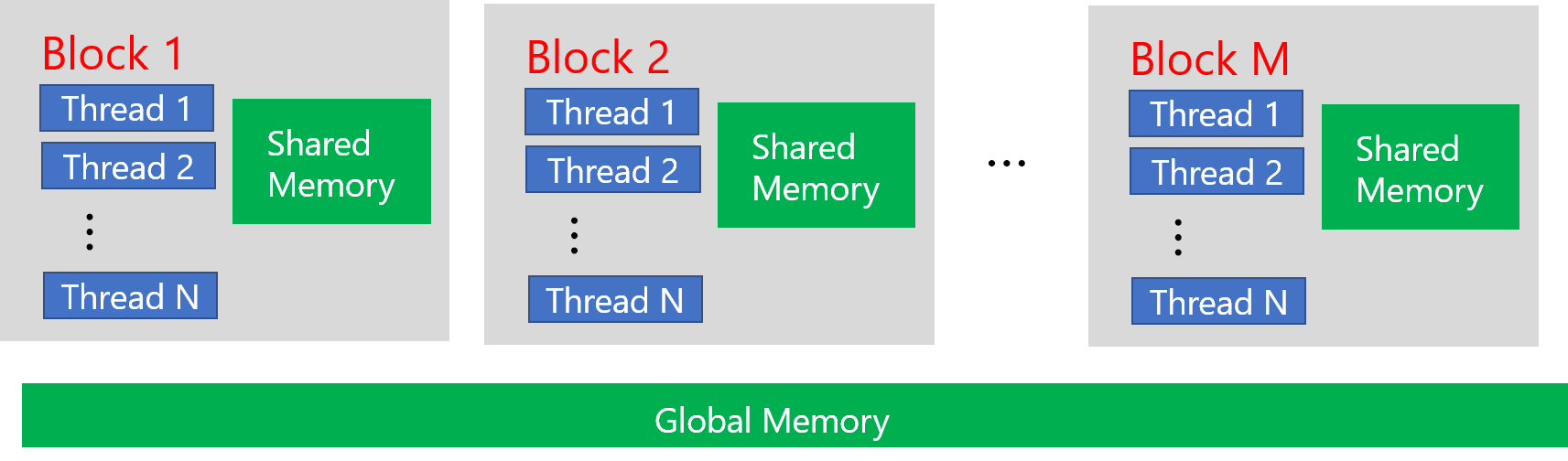}}
\caption{The structure of GPU from the viewpoint of software. }
\label{fig:GPU}
\end{figure}

\figurename~\ref{fig:GPU} shows the structure of a GPU from the viewpoint of software. A thread is a basic unit for parallel processing. Particle trackings are completely parallelized if each particle is assigned to a thread. A block is a group of threads. Each block has a shared memory to which only threads in the block can access. On the other hand, global memory can be accessed by any thread in the GPU. Although shared memory (48-96~kB) is smaller than global memory (16-32~GB) is, it has much higher bandwidth and lower latency. 

Nvidia provides a parallel computing platform called CUDA (Compute Unified Device Architecture)\cite{CUDA}. Using the CUDA platform, software developers can design applications executable by CUDA-enabled GPUs using programming languages such as C, C++, and Fortran. In fact, our new code is developed using C++.

\section{Single Particle Mechanics}
\label{sec:SPM}
Our developed code separately simulates particle motions by external fields and space charge effects.  To describe a single particle Hamiltonian, we use an independent variable $s$ which is the length along the reference orbit and 
a three-dimensional coordinate system $(x,y,\sigma)$, where $x$ and $y$ are two-dimensional coordinates on a plane perpendicular to the beam direction, and $\sigma$ is defined as $s-c\beta_{0}t$ using the velocity of the reference particle $\beta_{0}c$. Using this coordinate system with their conjugate variables 
$(p_{x},p_{y},p_{\sigma})$ and a vector potential $A_{s}(x,y)$, the Hamiltonian can be written as
\begin{equation}
\begin{split}
& H(x,p_{x},y,p_{y},\sigma,p_{\sigma}; s) \\
& = p_{\sigma} - (1+hx)\sqrt{(1+\delta)^{2}-p_{x}^{2}-p_{y}^{2}} -e\frac{A_{s}(x,y)}{p_{0}} \\
& \sim \frac{p_{x}^{2}+p_{y}^{2}}{2} + \frac{p_{\sigma}^{2}}{2\gamma_{0}^{2}} -hx-hxp_{\sigma} -\frac{p_{x}^{2}+p_{y}^{2}}{2}p_{\sigma}-e\frac{A_{s}(x,y)}{p_{0}} \\
& = H_{approx.}
\end{split}
\label{eq:H}
\end{equation}
where $h$, $\gamma_{0}$ and $p_{0}$ are the curvature, gamma factor and momentum of the reference particle, respectively. In additions, the momentum deviation $(p-p_{0})/p_{0}$ is expressed as $\delta$, which 
can be approximated as $p_{\sigma}+\frac{p_{\sigma}^2}{2\gamma_{0}^{2}}$.  How to solve the equation of motion depends on the types of external fields $A_{s}(x,y)$. Three different cases are shown as follows.
\subsection{Uniform Fields}
\begin{figure}[htbp]
\centerline{\includegraphics[width=2.5in]{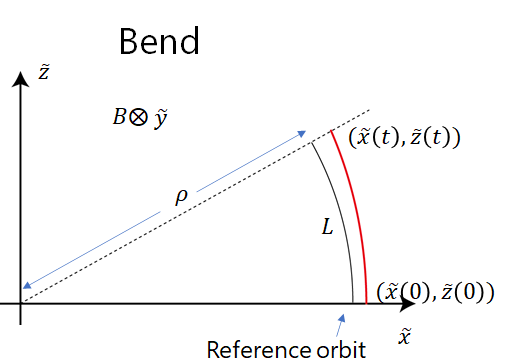}}
\caption{The Cartesian coordinate in uniform fields}
\label{fig:uniform}
\end{figure}
For uniform fields such as dipole magnets and drift spaces, the equation of motion ($\frac{d\vec{P(t)}}{dt}=e\vec{\beta}(t)c \times \vec{B}$, $t$ is time) can be analitically solved so that we need not use the approximate Hamiltonian ($H_{approx.}$ in (\ref{eq:H})). 
First we consider three-dimensional Cartesian coordinate $(\tilde{x},\tilde{y},\tilde{z})$ as shown in Fig~\ref{fig:uniform}. The directions of $\tilde{z}$ and $\tilde{y}$ are aligned to the reference orbit at the magnet entrance and the magnetid filed $\vec{B}=(0,B,0)$, respectively. The $\tilde{x}$ axis is choosen so that the system becomes left-handed. The origin is defined as the curvature center of the reference orbit. Since we consider only sector dipoles, we assume $\tilde{z}(0)=0$ as a initial condition. The solutions for $(\tilde{x},\tilde{z})$ as well as the corresponding momentums
($P_{\tilde{x}}(t), P_{\tilde{z}}(t)$) are obtained as 
\begin{equation}
\begin{split}
& \tilde{x}(t) = \frac{P_{\tilde{x}}(0)}{eB} \sin{\frac{eB}{m\gamma_{0}}t} + \frac{P_{\tilde{z}}(0)}{eB} \cos \frac{eB}{m\gamma_{0}}t + \tilde{x}(0)-\frac{P_{\tilde{z}(0)}}{eB} \\
& P_{\tilde{x}}(t) = \frac{P_{\tilde{x}}(0)}{eB} \cos \frac{eB}{m\gamma_{0}}t - P_{\tilde{z}}(0) \sin \frac{eB}{m\gamma_{0}}t \\
& \tilde{z}(t) = \frac{P_{\tilde{z}}(0)}{eB} \sin \frac{eB}{m\gamma_{0}}t - \frac{P_{\tilde{x}}(0)}{eB} \cos \frac{eB}{m\gamma_{0}}t + \frac{P_{\tilde{x}(0)}}{eB} \\
& P_{\tilde{z}}(t) = \frac{P_{\tilde{z}}(0)}{eB} \cos \frac{eB}{m\gamma_{0}}t + P_{\tilde{x}}(0) \sin \frac{eB}{m\gamma_{0}}t
\end{split}
\label{eq:BEND1}
\end{equation}
where, 
\begin{equation}
\begin{split}
& P_{\tilde{z}}(0) = p_{0}\sqrt{1+2p_{\sigma}+\beta_{0}^{2}p_{\sigma}^{2}-p_{x}(0)^2-p_{y}(0)^2} \\
& P_{\tilde{x}}(0) = p_{0}p_{x}(0).
\end{split}
\label{eq:BEND2}
\end{equation}
Here, the solution about the direction of the uniform field $B$ ($\tilde{y}$) is not shown since no force is applied to the direction. Secondly, the coordinate transformation from $(\tilde{x},\tilde{y},\tilde{z};t)$ 
to $(x,y,\sigma;s)$ is performed as
\begin{equation}
\begin{split}
& x(s=L) = \sqrt{\tilde{x}(t')+\tilde{z}(t')}-\rho \\
& p_{x}(s=L) = \frac{P_{\tilde{x}}(t')\cos\frac{L}{\rho}+ P_{\tilde{z}}(t')\sin\frac{L}{\rho}}{p_{0}} \\
& \sigma(s=L) = \sigma(0)  + L -\beta_{0}t', p_{\sigma} = Const.
\end{split}
\end{equation}
where the $t'$ can be obtained by the equation 
\begin{equation}
\tilde{z}(t') = \tilde{x}(t')\tan\frac{L}{\rho}
\end{equation}
as shown in \figurename~\ref{fig:uniform}
\subsection{Thick Quadrupole Magnets}
For thick quadrupole magnets ($eA_{s}(x,y)/p_{0} = -\frac{1}{2}k_{1}(x^{2}-y^{2})$), analytical solutions of the equation of motion for $H_{approx.}$ 
\begin{equation}
\begin{split}
 x(s) & = x(0)\cos\sqrt{k_{1}(1-p_{\sigma}(0))}s \\
& +\sqrt{\frac{1-p_{\sigma}(0)}{k_{1}}}p_{x}(0)\sin\sqrt{k_{1}(1-p_{\sigma}(0))}s \\
 y(s) & = y(0)\cosh\sqrt{k_{1}(1-p_{\sigma}(0))}s \\
& +\sqrt{\frac{1-p_{\sigma}(0)}{k_{1}}}p_{y}(0)\sinh\sqrt{k_{1}(1-p_{\sigma}(0))}s \\
p_{x}(s) & = p_{x}(0)\cos\sqrt{k_{1}(1-p_{\sigma}(0))}s \\
& -\sqrt{\frac{k_{1}}{1-p_{\sigma}(0)}}x(0)\sin\sqrt{k_{1}(1-p_{\sigma}(0))}s \\
 p_{y}(s) &= p_{y}(0)\cosh\sqrt{k_{1}(1-p_{\sigma}(0))}s \\
& +\sqrt{\frac{k_{1}}{1-p_{\sigma}(0)}}y(0)\sinh\sqrt{k_{1}(1-p_{\sigma}(0))}s \\
\sigma(s) &= -\frac{1}{2}\int_{0}^{s}ds'(p_{x}(s')^2+p_{y}(s')^2) + \frac{p_{\sigma}(0)}{\gamma_{0}^{2}}s \\
p_{\sigma}(s) &= p_{\sigma}(0) 
\end{split}
\end{equation}
are used since it is difficult to solve the exact equation of motion without any approximations.
The solution can be expressed as a symplectic transformation $e^{iH_{approx.}L}\vec{q_{0}}$ where $L$ and $\vec{q_{0}}$ are the length of the component along the reference orbit and the initial canonical variables, respectively. 
\subsection{Thick Sextupole Magnets}
For thick sextupole magnets ($eA_{s}(x,y)/p_{0} =  -\frac{1}{6}k_{2}(x^{3}-3xy^{2})$), analytical solutions are hardly obtained even for $H_{approx.}$. In this case, $H_{approx.}$ is divided into two parts as 
\begin{equation}
\begin{split}
H_{approx.} &=H_{0} + V \\
H_{0} &= \frac{p_{x}^{2}+p_{y}^{2}}{2} + \frac{p_{\sigma}^{2}}{2\gamma_{0}^{2}} - \frac{p_{x}^{2}+p_{y}^{2}}{2}p_{\sigma} \\
V & = \frac{1}{6}k_{2}(x^{3}-3xy^{2})
\end{split}
\end{equation}
where the equations for $H_{0}$ and $V$ are analytically solvable, then final state is obtained by multiple symplectic transformations described as  

\begin{equation}
e^{iH_{0}aL}e^{iVbL}e^{iH_{0}cL}e^{iVbL}e^{iH_{0}aL}
\label{eq:Sym2}
\end{equation}
where $a=\frac{1}{2}(1-\frac{1}{\sqrt{3}}), b = \frac{1}{2},  c=\frac{1}{\sqrt{3}}$\cite{Papaphilippou:2008zza}.

\section{Space Charge Effects}
\label{sec:SCE}
In this code, we assume the longitudinal length of a bunch is much larger than the transverse width. The assumption of long bunches is quite reasonable for the J-PARC Rapid Cycle Synchrotron\cite{Hotchi} and Main Ring\cite{Koseki}.
This corresponds to a two-dimensional approximation of the Poisson equation
\begin{equation}
\lambda(z)(\frac{\partial^{2}}{\partial x^{2}}+\frac{\partial^{2}}{\partial y^{2}})u(x,y)=-\frac{\lambda(z)f(x,y)}{\epsilon_{0}}.
\label{eq:poisson3D}
\end{equation}
Here, $\lambda(z)$ shows the line density of a bunch. For the numerical calculation shown in this section, the line density are created as $\sigma$-distributions with 128 bins where $\sigma$ ($=s-c\beta_{0}t$) can be obtained by solving the equation of motion for the Hamiltonian shown in Equation~\ref{eq:H}. The two-dimensional potential $u(x,y)$ includes the contribution from not only the charge and current density of a bunch themselves but their images through the beam pipes and magnetic poles. The electric fields can not penerate beam pipes (conductors) at any frequencies. On the other hand,  the magnetic fields only at low frequencies penerate the beam pipes so that the image current through the magnetic poles must be considered.   
Therefore, the effective three-dimensional potential $\phi(x,y,z)$ can be written as
\begin{equation}
\begin{split}
\phi(x,y,z) &=\frac{e}{m\gamma_{0}\beta_{0}^{2}c^{2}} \times \\
& ( (\lambda_{AC}(z)+\lambda_{DC})(u_{free}(x,y) + u_{image,\parallel}(x,y)) \\
& -\beta_{0}^{2}\lambda_{DC}(u_{free}(x,y)+u_{image,\perp}(x,y)) \\
& -\beta_{0}^{2}\lambda_{AC}(z)(u_{free}(x,y)+u_{image,\parallel}(x,y))).
\end{split}
\label{eq:pot3}
\end{equation}
Here, we divide the charge density $\lambda(z)$ into the DC part $\lambda_{DC}$ and the AC part $\lambda_{AC}(z)\equiv\lambda(z)-\lambda_{DC}$. $u_{free}$ is the electric potential in free space. $u_{image,\parallel}$ is the potential due the image charge for elimination of the electric field at the beam pipe. The coefficient $-\beta_{0}^{2}$ is used for the conversion from electric potentials to magnetic ones. $-\beta_{0}^{2}u_{image,\perp}(x,y)$ is the potential due to the image current which eliminates the tangential components of the magnetic field at the magnetic pole. 
Althouth our developed code involves the calculation of potentials of Equation~\ref{eq:pot3}, there is no experimetal or numerical benchmark so far.  Therefore, we adopt an additional approximation assuming the AC part of the bunch is larger than the DC, which means $\lambda(z)=\lambda_{AC}(z)$. This approximation is used for other tracking code called ``SCTR''\cite{Igarashi:HB2018-TUA2WD02}\cite{Ohmi} .  
As a result of the approximation, we obtaion  
\begin{equation}
\begin{split}
\phi(x,y,z) = &\frac{e}{m\gamma_{0}\beta_{0}^{2}c^{2}}(1-\beta_{0}^{2})\times \\ 
&\lambda(z)(u_{free}(x,y) + u_{image,\parallel}(x,y)) \\
=&\frac{e}{m\gamma_{0}^{3}\beta_{0}^{2}c^{2}}\lambda(z)(u_{free}(x,y) + u_{image,\parallel}(x,y)) 
\end{split}
\label{eq:pot2}
\end{equation}
In this approximation, we just have to solve the two-dimensional Poisson equation shown in Equation~\ref{eq:poisson3D} with the boundary condition $u(x,y) = 0$. 
The transverse kick due to space charge effects is calculated as  
\begin{equation}
 -\frac{e}{m\gamma_{0}^{3}\beta_{0}^{2}c^{2}}\lambda(z)(\frac{\partial}{\partial x}, \frac{\partial}{\partial y})u(x,y) \times L 
\end{equation}
where $L$ is the distance from the previous location at which space charge effects are calculated.
The longitudinal kick due to space charge effects is calculated as 
\begin{equation}
 -\frac{e}{m\gamma_{0}^{2}\beta_{0}^{2}c^{2}}u(x,y) \frac{\partial}{\partial z}\lambda(z)\times L
\end{equation}
In this section, we describe how to make two-dimensional charge densities $f(x,y)$ and solve two-dimensiol Poisson equiations for the potential u(x,y) $u(x,y)$. 
Although this code enables us to choose Cartesian $f(x,y)$ or polar $f(r,\theta)$ coordinates depending on the cross-sections of beam pipes, the descriptions in this section are based on the Cartesian coordinate.
 \subsection{Charge Density Calculation}
 \label{sec:rho}
\begin{figure}[htbp]
\centerline{\includegraphics[width=2.0in]{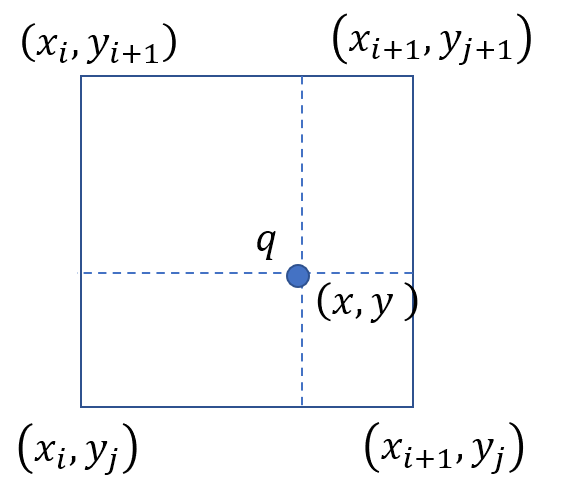}}
\caption{The example of a relation between a particle location and the adjacent grids }
\label{fig:ChargeWeight}
\end{figure}

A two-dimensional charge density $f(x,y)$ fits the size of shared memory of GPUs when the number of cells is approximately about 10000 (100$\times$100). 
In fact, a state-of-the-Art GPU can allocate 96 kB shared memory, which corresponds to 12000 double-precision floating-point numbers. 
Charge densities are calculated using the final states of all particles for each component. When a particle located at $(x,y)$ is in the rectangle whose vertices are four grids labeled as $(x_{i},y_{j})$, $(x_{i+1},y_{j})$, $(x_{i},y_{j+1})$ and $(x_{i+1},y_{j+1})$ (\figurename~\ref{fig:ChargeWeight}), 
a two-dimensional histogram is filled as
\begin{equation}
\begin{split}
& Q(x_{i},y_{j}) += \frac{(x_{i+1}-x)(y_{j+1}-y)}{\delta x \delta y} \\
& Q(x_{i+1},y_{j}) += \frac{(x-x_{i})(y_{j+1}-y)}{\delta x \delta y} \\
& Q(x_{i},y_{j+1}) += \frac{(x_{i+1}-x)(y- y_{j})}{\delta x \delta y} \\
& Q(x_{i+1},y_{j+1}) += \frac{(x-x_{i})(y - y_{j})}{\delta x \delta y} 
\end{split}
\end{equation}
where, $\delta x = x_{i+1}-x_{i}, \delta y = y_{j+1}-y_{j}$.
\begin{figure}[htbp]
\centerline{\includegraphics[width=3.5in]{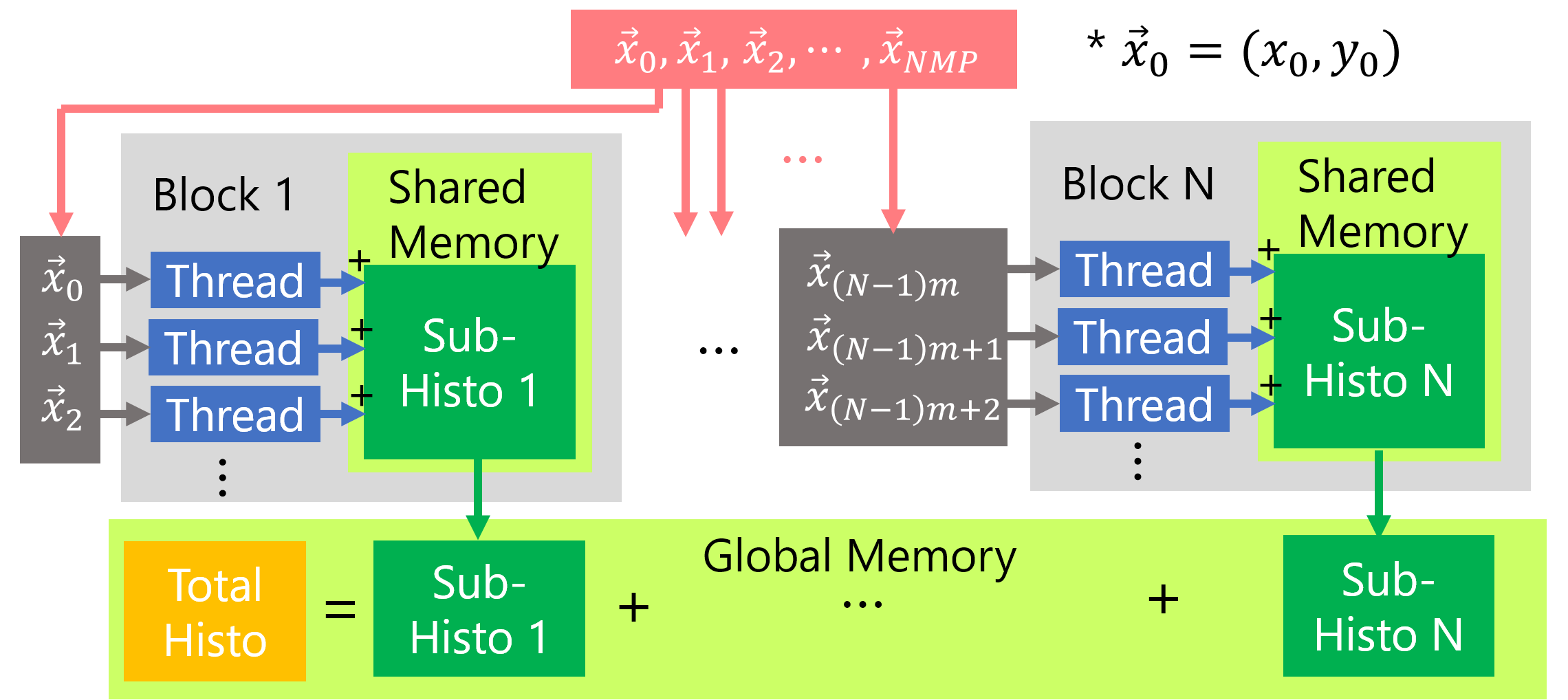}}
\caption{Sub-histograms for the reduction of colliding threads.}
\label{fig:Subhisto}
\end{figure}
The entries and bins correspond to the charges and grids, respectively.  In case that a thread fills the histogram for a single particle, collisions between threads occur when filling a common bin of the histogram. This is why not all threads can be executed in parallel. To reduce these colliding threads, as shown in \figurename~\ref{fig:Subhisto}, all threads in a block fill a sub-histogram allocated at shared memory of the block so that fewer threads fill a common histogram\cite{CUDABOOK}. Once all sub-histograms are filled, they are summed. It must be noted that colliding threads can still occur, but much less frequently. We thus use a special operation called an atomic operation provided by the CUDA platform to fill the sub-histograms. When an atomic operation accesses data at some address, other memory accesses to the same address are blocked until the operation is done. 

\subsection{Poisson Solver}
Using the charge distribution $f(x,y)$, potential $u(x,y)$ is obtained by solving two-dimensional Poisson equation 
\label{sec:phi}
\begin{equation}
(\frac{\partial^{2}}{\partial x^{2}}+\frac{\partial^{2}}{\partial y^{2}})u(x,y)=f(x,y)
\label{eq:poisson2D}
\end{equation}
with the boundary conditions 
\begin{equation}
u(x,0) = u(L_{x},y) = u(0,y) = u(x,L_{y}) = 0
\label{eq:boundary}
\end{equation}
where $L_{x}$ and $L_{y}$ are the horizontal and vertical length of a beam pipe.
The boundary conditions suppress the electric fields at the inner surface of the beam pipe.
The CUDA platform involves a library of FFT (Fast Fourier Transform) called ``cuFFT''\cite{cuFFT}.
The functions in cuFFT are designed to provide high performance on CUDA-enabled GPUs. We thus employ DFTs (Discrete Fourier Transform) for the Poisson solver.
\newpage
The differential equation (\ref{eq:poisson2D}) is discretized as 
\begin{equation} 
\begin{split}
& \frac{u_{i-1,j}-2u_{i,j}+u_{i+1,j}}{\delta x^2} + \frac{u_{i,j-1}-2u_{i,j}+u_{i,j+1}}{\delta y^2} =  f_{i,j} \\
& i,j = 1,2,\ldots m
\end{split}
\label{eq:disc}
\end{equation}
where $m$ is the number of cells in one direction. The odd extensions of $u_{i,j}$ and $f_{i,j}$, which are labeled as $V$ and $F$, are constructed as 
\begin{strip}
\begin{equation}
V \equiv \left(
    \begin{array}{cccccccccc}
      0 & 0 & 0 & \ldots & 0 & 0 & 0 & 0 & \ldots & 0 \\
      0 & u_{1,1} & u_{1,2} & \ldots & u_{1,m} & 0 & -u_{1,m} &  -u_{1,m-1}  & \ldots & -u_{1,1} \\
      0 & u_{2,1} & u_{2,2} & \ldots & u_{2,m} & 0 & -u_{2,m} &  -u_{2,m-1}  & \ldots & -u_{2,1} \\   
      \vdots & \vdots & \vdots & \ddots & \vdots & \vdots & \vdots &  \vdots & \ddots & \vdots \\      
      0 & u_{m,1} & u_{m,2} & \ldots & u_{m,m} & 0 & -u_{m,m} &  -u_{m,m-1}  & \ldots & -u_{m,1} \\  
      0 & 0 & 0 & \ldots & 0 & 0 & 0 & 0 & \ldots & 0 \\
        0 & -u_{m,1} & -u_{m,2} & \ldots & -u_{m,m} & 0 & u_{m,m} &  u_{m,m-1}  & \ldots & u_{m,1} \\  
            0 & -u_{m-1,1} & -u_{m-1,2} & \ldots & -u_{m-1,m} & 0 & u_{m-1,m} &  u_{m-1,m-1}  & \ldots & u_{m-1,1} \\
               \vdots & \vdots & \vdots & \ddots & \vdots & \vdots & \vdots &  \vdots & \ddots & \vdots \\    
               0 & -u_{1,1} & -u_{1,2} & \ldots & -u_{1,m} & 0 & u_{1,m} &  u_{1,m-1}  & \ldots & u_{1,1} \\  
    \end{array}
  \right)
  \label{eq:extension}
\end{equation}
\end{strip}
where only $V$ is shown but the $F$ can be constructed in the same manner.
The equation (\ref{eq:disc}) of the $V$ and $F$ instread of the $u$ and $f$ 
\begin{equation} 
\begin{split}
& \frac{V_{l-1,l'}-2V_{l,l'}+V_{l+1,l'}}{\delta x^2} + \frac{V_{l,l'-1}-2V_{l,l'}+V_{l,l'+1}}{\delta y^2} =  F_{l,l'} \\
& l,l' = 0,1,2,\ldots 2m+1
\end{split}
\label{eq:discextension}
\end{equation}
is also satisfied. Using the $V$, the boundary conditions (\ref{eq:boundary}) become   
\begin{equation}
 V_{l,m+1}=V_{l,0}=V_{0,l'}=V_{m+1,l'}=0
\end{equation}
which are satisfied by definition of (\ref{eq:extension}).

\begin{figure}[htbp]
\centerline{\includegraphics[width=3.0in]{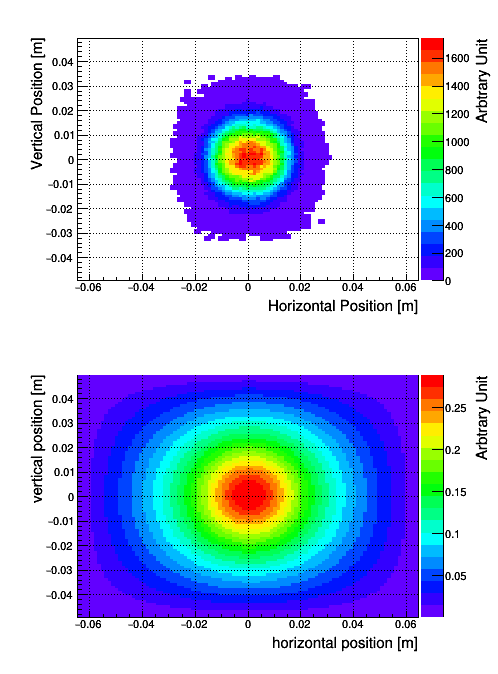}}
\caption{The upper and lower figures are a two-dimensional charge density and the potential, respectively. }
\label{fig:rhopot}
\end{figure}

One-dimensional DFT is defined as 
\begin{equation}
\begin{split}
 & DFT_{l}(g_{l})_{p} \equiv \sum_{l=0}^{N-1}g_{l}e^{-i\frac{2\pi pl}{N}} \\
 & p = 1,2,\ldots,N 
\end{split} 
\label{eq:DFT}
\end{equation}
where $l$ is an index to one direction. Applying DFT about one direction labeled as $l$
to the first term of the left-hand side in (\ref{eq:discextension}), we obtain
\begin{strip}
\begin{equation}
\begin{split}
& DFT_{l}(V_{l-1,l'}-2V_{l,l'}+V_{l+1,l'})_{p} = u_{1,l'} + e^{-i\frac{p}{2(m+1)}2\pi}(-2u_{1,l'}+u_{2,l'}) + e^{-i\frac{2p}{2(m+1)}2\pi}(u_{1,l'}-2u_{2,l'}+u_{3,l'}) \cdots \\
& = 0 + (1-2e^{-i\frac{p}{2(m+1)}2\pi}+e^{-i\frac{2p}{2(m+1)}2\pi})u_{1,l'} + (e^{-i\frac{p}{2(m+1)}2\pi}-2e^{-i\frac{2p}{2(m+1)}2\pi}+e^{-i\frac{3p}{2(m+1)}2\pi})u_{2,l'} \cdots \\
& =  (e^{-i\frac{p}{2(m+1)}2\pi} +e^{i\frac{p}{2(m+1)}2\pi} - 2 )(0+e^{-i\frac{p}{2(m+1)}2\pi}u_{1,l'} + e^{-i\frac{2p}{2(m+1)}2\pi}u_{2,l'} \cdots \\
& = -4 \sin^2\frac{p\pi}{2(m+1)}\sum_{l=0}^{2(m+1)-1}e^{-i\frac{pl}{2(m+1)}2\pi}V_{l,l'}.
\end{split}
\label{eq:DFT1}
\end{equation}
\end{strip}
Another DFT to ($\ref{eq:DFT1}$) about the other direction labeled as $l'$ gives
\begin{equation}
\begin{split}
& DFT_{l'}(DFT_{l}(V_{l-1,l'}-2V_{l,l'}+V_{l+1,l'})_{p})_{q} \\
& = -4 \sin^2\frac{p\pi}{2(m+1)}\sum_{l=0}^{2(m+1)-1}\sum_{l'=0}^{2(m+1)-1}e^{-i\frac{(pl+ql')\pi}{(m+1)}}V_{l,l'} \\
& = -4 \sin^2\frac{p\pi}{2(m+1)}DFT_{l'}(DFT_{l}(V_{l,l'})_{p})_{q}.
\end{split}
\end{equation}
By adding the DFT to the second term of the left-hand side in (\ref{eq:discextension}), total two-dimensional DFT of 
the left-hand side in (\ref{eq:discextension}) is written as
\begin{equation}
\begin{split}
& DFT_{l'}(DFT_{l}(\frac{V_{l-1,l'}-2V_{l,l'}+V_{l+1,l'}}{\delta x^2})_{p})_{q} \\
& + DFT_{l'}(DFT_{l}(\frac{V_{l,l'-1}-2V_{l,l'}+V_{l,l'+1}}{\delta y^2})_{p})_{q} \\
& = -4 (\frac{1}{\delta x^2}\sin^2\frac{p\pi}{2(m+1)}+\frac{1}{\delta y^2}\sin^2\frac{q\pi}{2(m+1)}) \\
& \times DFT_{l'}(DFT_{l}(V_{l,l'})_{p})_{q}.
\end{split}
\end{equation}
Since this equals to the two-dimensional DFT of $F_{l,l'}$, the two-dimensional DFT of $V_{l,l'}$ is derived as
\begin{equation}
\begin{split}
& DFT_{l'}(DFT_{l}(V_{l,l'})_{p})_{q}  \\
& = - \frac{DFT_{l'}(DFT_{l}(F_{l,l'})_{p})_{q} }{4 (\frac{1}{\delta x^2}\sin^2\frac{p\pi}{2(m+1)}+\frac{1}{\delta y^2}\sin^2\frac{q\pi}{2(m+1)})} 
\end{split}
\end{equation}
Using inverse DFT defined as 
\begin{equation}
iDFT_{p}(G_{p})_{l} \equiv \frac{1}{N}\sum_{p=0}^{N-1}G_{p}e^{i\frac{2\pi pl}{N}}
\end{equation}
The potential $V_{l,l'}$ can be obtained as  
\begin{equation}
V_{l,l'}  = - \frac{iDFT_{q}(iDFT_{p}(DFT_{l'}(DFT_{l}(F_{l,l'})_{p})_{q})_{l})_{l'}}{4 (\frac{1}{\delta x^2}\sin^2\frac{p\pi}{2(m+1)}+\frac{1}{\delta y^2}\sin^2\frac{q\pi}{2(m+1)})}. 
\label{eq:solution}
\end{equation}
This solution (\ref{eq:solution}) involves two DFT and two inverse DFT operations, which are calculated using the functions provided by cuFFT. The \figurename~\ref{fig:rhopot} shows a two-dimensional charge density and the corresponding potential.  The area in both two-demensional histograms corresponds to the cross section of a beam pipe. This is 
the reason why the potential becomes constant at the boundary.  
\subsection{Interpolation}
\begin{figure}[htbp]
\centerline{\includegraphics[width=2.5in]{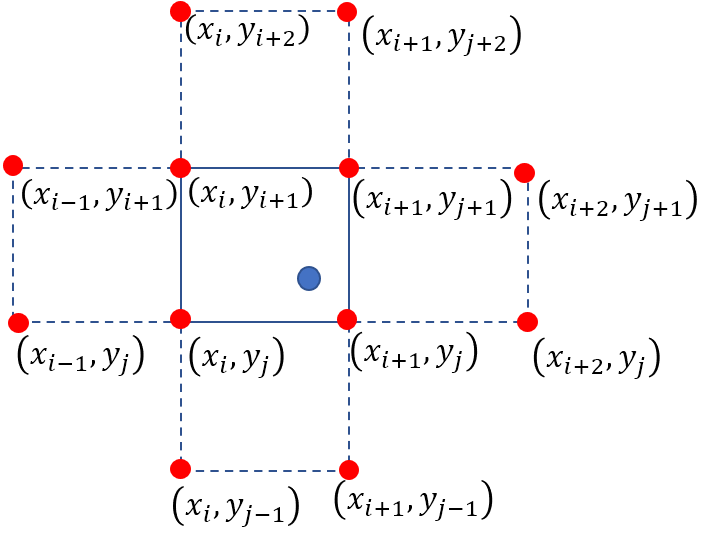}}
\caption{The example of a relation between a particle location and the adjacent grids. }
\label{fig:Interpolation}
\end{figure}
The potential at $(x,y)$ as shown in \figurename~\ref{fig:Interpolation} is obtained by the interpolation using a Bezier surface written as
\begin{equation}
\begin{split}
 u (x,y) & = \sum_{m=0}^{3}\sum_{n=0}^{3} ( u_{i+m-1,j+n-1} \\
& \times \frac{3!}{m!(3-m)!}(\frac{x-x_{i}}{\delta x})^m (\frac{x_{i+1}-x}{\delta x})^{3-m} \\
& \times \frac{3!}{n!(3-n)!}(\frac{y-y_{j}}{\delta y})^n (\frac{y_{j+1}-y}{\delta y})^{3-n} ).
\end{split}
\end{equation}
The electric field can be obtained as
\begin{equation}
\vec{E}(x,y) = -\nabla u(x,y) = - (\frac{\partial}{\partial x},\frac{\partial}{\partial y}) u(x,y)
\end{equation} 

\section{The J-PARC Main Ring as a Simulation Example}
\label{sec:JPARC}
\begin{figure}[htbp]
\centerline{\includegraphics[width=3.0in]{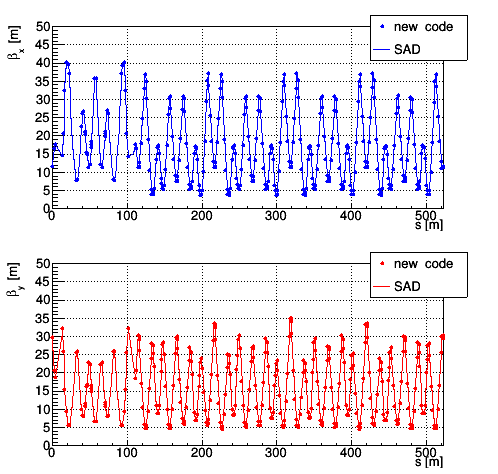}}
\caption{The betatron amplitude functions of the J-PARC Main Ring.}
\label{fig:Beta}
\end{figure}

\begin{figure}[htbp]
\centerline{\includegraphics[width=3.0in]{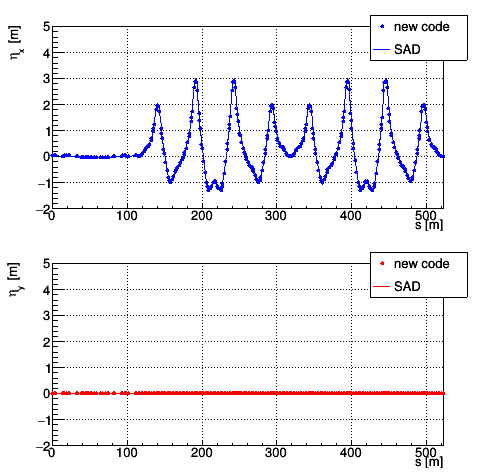}}
\caption{The dispersion functions of the J-PARC Main Ring. }
\label{fig:Disp}
\end{figure}

\begin{figure}[htbp]
\centerline{\includegraphics[width=3.0in]{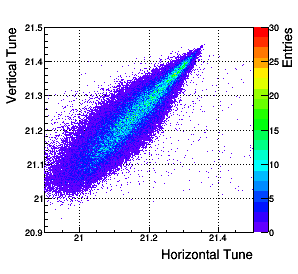}}
\caption{The two-dimensional distribution of the betatron tunes of all tracked particles (our developed code)}
\label{fig:Tune}
\centerline{\includegraphics[width=3.0in]{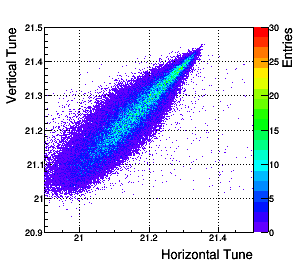}}
\caption{The two-dimensional distribution of the betatron tunes of all tracked particles (SCTR)}
\label{fig:TuneSCTR}
\end{figure}

To verify the developed code, we simulated several parameters of the J-PARC Main Ring. 
Table~\ref{tab100} shows the main parameters of the J-PARC Main Ring\cite{Koseki}. The simulation parameters are shown in Table~\ref{tab200}. The components involve drift spaces, RF caviies and short corrector magnets as well as the magnets shown in Table~\ref{tab100}. The locations of the space charge calculation are chosen so that 
their intervals are approximately 1~m or less. For example, the calculation in a bending magnet, whose length approximately 6~m, is performed 6 times. We fixed the intervals to reduce the beam losses due to numerical errors (unphysical beam losses without any imperfections ) down to much smaller than actually observed. This is very important tasks because the code should be useful for actual beam tunings. The transverse beam emittance, momentum spread and bunch factor are based on the measurements. The bunch factor is increased up to 0.3 during the ring circulation due to second harmonic RF technique, which is also implemented in the code. For the chromatic correction, the sextuple strength is decreased by 25 \% comparing to the full correction.      
\begin{table}[htbp]
\caption{The main parameters of the J-PARC Main Ring}
\label{table100}
\centering
\begin{tabular}{cc}
\hline
Cycle time [s] & 2.48 or 5.2 \\
\hline
Injection energy [GeV] & 3 \\
\hline
Extraction energy [GeV] & 30 \\
\hline
Harmonic number (h) & 9 \\
\hline
Number of bunches & 8 \\
\hline
Number of Protons per bunch ($N_{ppb}$) & $3.4 \times10^{13}$ \\
\hline
Physical aperture [$\pi$mm-mrad] & 81 \\
\hline
Collimator aperture [$\pi$mm-mrad] & $\sim$60 \\
\hline
Number of bending magnets & 96 \\
\hline
Number of quadrupole magnets & 216 \\
\hline
Number of sextupole magnets & 72 \\
\hline
\end{tabular}
\label{tab100}
\end{table}
\begin{table}[htbp]
\caption{The main parameters of the simulation for the J-PARC Main Ring}
\label{table200}
\centering
\begin{tabular}{cc}
\hline
Number of macro-particles & 200000 \\
\hline
Number of components per turn & 3717 \\
\hline
Number of calculations of space charge effects per turn & 2109 \\
\hline
GPU & TESLA-V100 \\
\hline
Double precision floating point operations [1/s] & $7\times10^{12}$ \\
\hline
Initial 1$\sigma$ transverse beam emittance ($\epsilon_{1\sigma}$) [$\pi$mm-mrad] & 4 \\
\hline
Initial momentum spread (full width) [\%] & 0.2 \\
\hline 
Initial bunch factor ($B_{f}$)& 0.2 \\
\hline
\end{tabular}
\label{tab200}
\end{table}
The betatron amplitude functions and dispersion functions obtained by the code are compared to the calculation by SAD (Strategic Accelerator Design) \cite{SAD}. The SAD calculations employ transfer matrics. On the other hand, the new code calculates them in different ways. Assuming that the transverse particle distribution is gaussian, the betatron amplitude function at $s$ ($\beta(s)$) can be written as $\sigma^2(s)/\epsilon_{1\sigma}$ where $\sigma(s)$ is the standard deviation of the transverse distribution at $s$, and $\epsilon_{1\sigma}$ is the 1$\sigma$ emittance. For the new code, the $\beta(s)$ are obtained by calculating the RMS of the transverse coordinates ($x$ or $y$) of all tracked particles without space charge effects. For the dispersion functions in the new code, we use the central orbit distortions of off-momentum particles, which are actually tracked.  As shown in \figurename~\ref{fig:Beta} and \ref{fig:Disp}, the results from the new code reproduce the SAD calculation. These validate single-particle mechanics in the new code described in Section~\ref{sec:SPM}.  
To validate space charge effects in the new code, we calculate the betatron tune shift of all tracked particles.
\figurename~\ref{fig:Tune} shows the betatron tunes of all tracked particles. The tunes are obtained as the transverse phase advances througth one turn. 
The same plot using ``SCTR'''\cite{Igarashi:HB2018-TUA2WD02}\cite{Ohmi} is also shown in \figurename~\ref{fig:TuneSCTR}. Our developed code clearly reproduces the result from ``SCTR''. The betatron tunes without space charge effects (only due to the strength of the quadrupole magnets) are set at $(\nu_x,\nu_y)=$(21.35,21.44).
The estimated tune spread $\Delta\nu_{x,y}$ can be obtained as 
\begin{equation}
\Delta\nu_{x,y} = - \frac{N_{ppb}hr_{p}}{4\pi\beta_{0}^{2}\gamma_{0}^{3}\epsilon_{1\sigma} B_{f}} = -0.45. 
\label{eq:tuneshift}
\end{equation}
As shown in Table~\ref{tab100}, we use $3.4\times10^{13}$, 9, 4 for $N_{ppb}$, $h$ and $\epsilon_{1\sigma}$, respectively. For the bunch factor, we use 0.3 for Equation~\ref{eq:tuneshift}, \figurename~\ref{fig:Tune} and \figurename~\ref{fig:TuneSCTR}. $r_p$ is classical proton radius ($\approx1.547\times10^{-18}$~[m]). The numerical result shown in \figurename~\ref{fig:Tune} is consistent with the rough estimation using the formula.

The number of turns of the J-PARC Main Ring which can be simulated per minute is used as a benchmark. For this test, the same numbers of macro-particles (200000) are used for both codes. The potential calculation are different between two. Our new code sets a potential boundary on the inner surface of a beam pipe and  
selects a polar or rectangular coordinate depeding on the cross section of the beam pipe. Due to the limitation of shared memories, we adopt a 100$\times$100 grid for both coordinates. On the other hand, SCTR uses a common rectangle boundary and use a 128$\times$128 grid for all locations in the ring.  
Using the new code on a single TESLA-V100 GPU, we obtained approximately 117 turns per minute in contrast to 9 for SCTR\cite{Ohmi} on Intel Xeon(R)Gold 6126 (2.6GHz). This is a significant difference. In fact, the beam losses in the J-PARC main ring are distributed from the injection until approximately $5\times10^5$th turn.  

\section{Summary}
\label{sec:summary}
Particle tracking simulations including space charge effects are very important for high-intensity proton rings. Since they include not only
Hamilton mechanics of a single particle but constructing charge densities and solving Poisson equations to obtain the electromagnetic
field due to the space charge, they are extremely time-consuming. We have newly developed a particle tracking simulation code that can be
used in GPU. 

GPUs have strong capacities of parallel processing so that the calculation of single-particle
mechanics can be done very fast by complete parallelization. Our new code also includes the space charge effect. It must construct charge
densities, which cannot be completely parallelized. For the charge density construction, we fill sub-histograms in shared memory before constructing 
the total histogram so that each thread can not only avoid their frequent collisions but access sub-histograms very fast.  
For the Poisson solver, we employ DFT to take advantage of the usage of the cuFFT library, which is designed to provide high performance on CUDA-enabled GPUs.

To validate single-particle mechanics and space charge effects implemented in the code, the betatron amplitude functions, dispersion functions, and space-charge-induced tune spread are simulated in the  case of the J-PARC Main Ring. These results reproduce the calculations by other simulators. In addition, the new code on a single TESLA-V100 GPU  can simulate approximately 117 turns per minute in contrast to 9 for SCTR\cite{Ohmi} on Intel Xeon(R)Gold 6126 (2.6GHz).

\end{document}